# A flapping feathered wing-powered aerial vehicle

Zhenhong Zhang[1], Aiqiu Wei[2], Yu Cai[3]*

**An aerial vehicle powered by flapping feathered wings was designed, developed and fabricated. Different from legacy flapping-wing aerial vehicles with membrane wings, the new design uses authentic bird feathers to fabricate wings. In field tests, a radio-controlled electric-powered aerial vehicle with flapping feathered wings successfully took off, flew up to 63.88 s and landed safely. It was found that flapping feathered wings can generate sufficient thrust and lift to make a man-made aerial vehicle accomplish takeoff, sustainable flight and a safe landing.**

## INTRODUCTION

In 1874, Alphonse Penaud *(1)* developed the first rubber-band-powered flyable flapping-wing aerial vehicle. Later, a commercial product, named Tim Bird *(2)*, powered by rubber bands, on which many of the following ornithopter designs are based, was released to the market. In recent years, many bionic flapping-wing robotic platforms inspired by flying creatures have been developed *(3)*. Among them, Robofly *(4,5)* and Delfly *(3,6)* are well-known examples of insect-level flapping-wing aerial vehicles. Birds, the most popular flying animals, attract more attention than other creatures *(7-9)*. Many successful flyable bird-like flapping-wing aerial vehicles have been invented; for example, Slow Hawk *(10)* uses a radio-controlled (RC) motor and transmission system to control the synchronously flapping of wings, which generate aerodynamic power to make the robotic bird fly. In subsequent research, many ornithopters with structures similar to that of Slow Hawk were developed, among which Phoenix *(11)* is a typical example. Another typical ornithopter is Robo Raven *(12)*. Unlike Slow Hawk, Robo Raven uses two independent RC servo motors to independently control the flapping of its wings. For Slow Hawk, the flapping of the two wings is synchronized, which means the flapping is in-phase, in-rate and interdependent *(10)*. The wings of either Slow Hawk or Robo Raven are fabricated with carbon fiber skeletons covered with a plastic foam membrane or thin plastic film *(13,14)*. In 2011, an ornithopter named Smartbird was developed by Festo *(15-17)* and successfully flew. Smartbird has a novel hinged-wing design. Each wing consists of two sections: the inner wing and outer wing. The inner wing is connected to the outer wing by hinges *(15)*. For each wing, there are two or three meticulously designed four-bar linkage systems; the power is transmitted through the torso to the two wing sections, making each wing section flap with different kinematics. The different movements of the inner and outer wings makes Smartbird extremely authentic. Although its wings look like 2-degree-of-freedom (DOF) robotic bird wings, the movements of the inner wing and outer wing are not independently driven, and their movements are interrelated so that the overall DOF is less than 2 *(15)*. Either the inner wing or outer wing of Smartbird is made of membrane-like materials, including extruded polyurethane foam *(17)*, which is different from a bird's feathered-wing structure. In 2020, Lentink's team developed a flyable aerial vehicle called PigeonBot *(19)*, whose wings are made with real bird feathers. This demonstrated that the asymmetric morphing of wings and different feather positions can control the flight direction. Unlike birds powered by flapping wings, PigeonBot is powered by a propeller. Creating a flyable flapping-wing robot with feathered and deformable wings has always been an open question *(19,20)*. This research implemented the idea of implanting real bird feathers on a hinged-wing robot and utilizing flapping feathers to generate aerodynamic power. The robot powered by flapping feathered wings successfully took off and flew for more than 60 s and then ultimately land safely. Field tests proved that carefully designed flapping feathers can generate significant aerodynamic power to drive man-made aerial vehicles.

[1]School of computer, electronic & information engineering, Guangxi University; Nanning, GuangXi, China.
[2]Bee-eater Technology Inc.; Nanning, GuangXi, China.
[3]School of electrical engineering, Guangxi University; Nanning, GuangXi, China.
*Corresponding author. Email:ramanujancn@hotmail.com

## WINDRIDER TEST PLATFORM

To ensure successful takeoff and sustainable flight, the structure of an aerial vehicle has to be lightweight and strong. The authors compared various materials and finally selected a 1.5 mm thick carbon fiber sheet to build torso frames and wing ribs. The backbone of the torso is made of a 6 mm carbon fiber rod, and the backbone of the wing is made of a 4 mm carbon fiber rod. Parts of the load-bearing points and joints are made with 6063 aluminum alloy. The regular parts were made with a 3D printer. The printing material was polylactic acid (PLA). More details are shown in Fig. 1.

The designed robot is named WindRider, and its design is based on that of Smartbird. The authors designed a carbon fiber frame for the main transverse frame of the robot. The motor, battery, transmission gear and four-bar linkages are housed on the main transverse frame and torso backbone. The running motor drives the gearbox with a 1:48 reduction ratio. After speed reduction, the output of the gearbox drives the wings up and down via a four-bar linkage system (Fig. 1E). Essentially, there are two four-bar linkages in the transmission system, one driving the inner wing section and the other driving the outer wing section. The second four-bar linkage couples upon the first four-bar linkage; therefore, they are interrelated. The inner wing generates lift, and the outer wing generates thrust *(15)*. A 3 mm thick foaming rate of 45 expanded polypropylene (EPP) film attached to the tail frame by foam glue *(18)* was used as the vertical stabilizer. The tail of the aerial vehicle is controlled by RC transceiver systems WFT07 and WFR07S *(24)* to accomplish the pitch and yaw movements of the robot (Fig. 1D). When the tail rotates around the elevator axis (Fig. 1B), a pitch moment is generated (Fig. 1D) to adjust the pitch angle of the robot. When the tail rotates around the rudder axis (Fig. 1B), a yaw moment is generated around the yaw axis of the robot (Fig. 1D), causing the vehicle to turn left or right *(17)*. According to the field experience of the controller of WindRider, who is also one of authors, qualitatively, the yaw controlling torque can be partly coupled with the roll controlling portion. Because coupling effects are insignificant, it is difficult for robots to make sharp turns or other acrobatic movements. More quantitative research on this phenomenon is required in future studies. To protect the torso frame and reduce aerodynamic drag, a streamline seagull-like EPP outer shell (Fig. 1C) was fabricated. The torso frames and outer shell are adhered with foam glue. The final version of WindRider's skeleton is shown in Fig. 1 E to F.

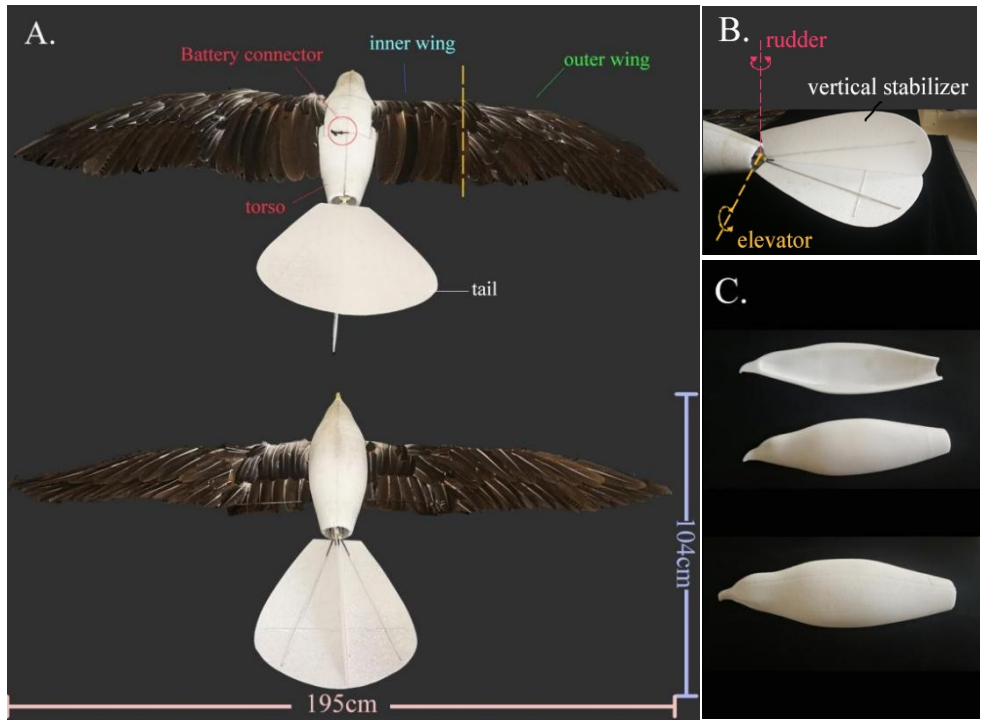

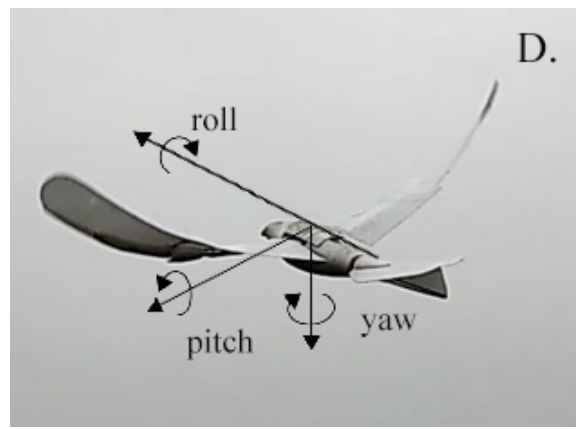

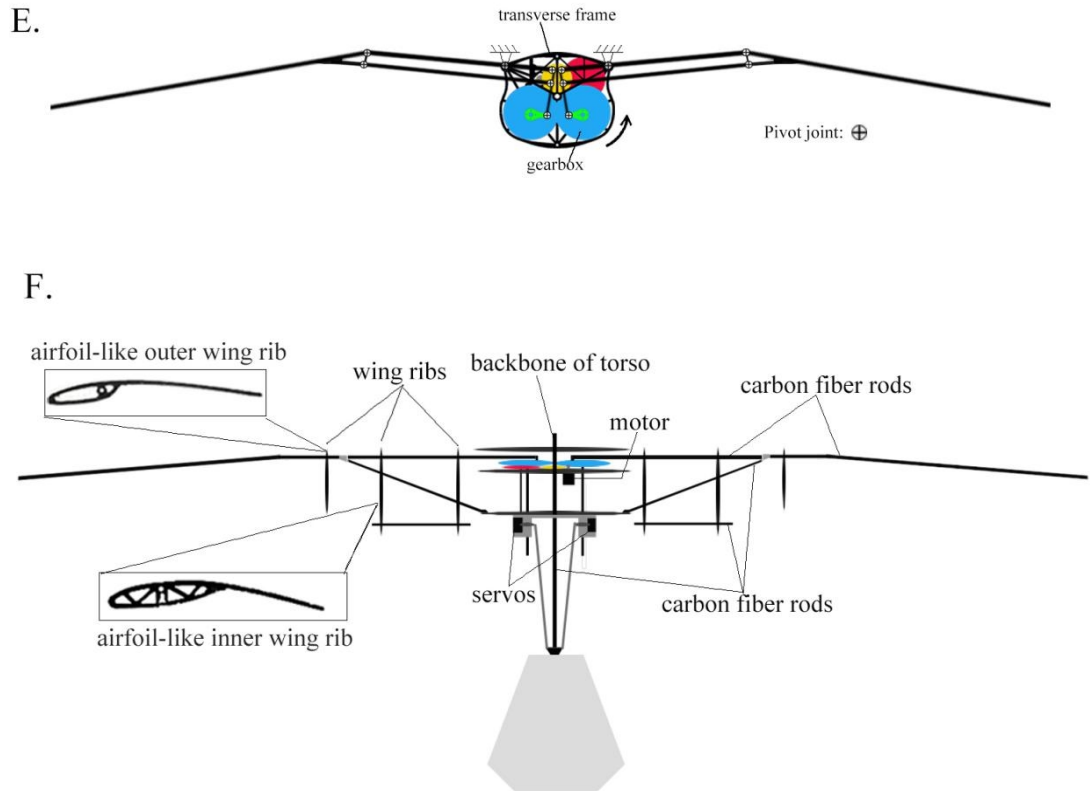

**Fig. 1. WindRider: A bionic flapping-wing aerial vehicle with feathered wings that consists of a torso, a tail and hinged feathered wings.** (A). Top and bottom views of WindRider. It has a wingspan of 195 cm, an inner wingspan of 32 cm, an outer wingspan of 65 cm, a length (from nose to tail) of 104 cm, an overall weight of 667.2 g (including a 3-cell Li-po battery with a capacity of 350 mAh and rated voltage of 11.7 V), and a takeoff flapping frequency of ~ 4 Hz. (B). The tail has a vertical stabilizer, and the vertical stabilizer is attached to the tail frame to make vehicle left or right turn. (C). The seagull-like torso is made of EPP and carved by CNCs (computer numerical control). (D). The 3 axes of the robot are roll, pitch, and yaw. (E). The front view of the skeleton and the transverse frame of the torso, cranks, and wing ribs are made of carbon fiber sheets carved by a CNC machine. The gearbox is made of 6063 aluminum alloy. (F). The top view of the skeleton. The tail and the torso are connected by a 3D-printed plastic part. The maximum continuous power of the motor (Sunnysky-2206,1900 kV) is 150 W. Two servos (EMAX ES09MD with a torque of 2.3/2.6 kg.cm and a speed of 0.10/0.08 s/60°) are used to control the tail during yaw and pitch movements. The wing ribs are airfoil-like shapes supported by 3 mm and 4 mm carbon fiber rods. The inner wing ribs and outer wing ribs have different shapes, as shown in Fig. 1F.

## FIELD TESTS AND PREPARATIONS

The authors designed and developed a reusable wing with a carbon fiber skeleton and authentic bird feathers for the covering material. Before making this complicated feathered wing, the authors ordered a large number of feathered fans from Taobao.com *(23)* and disassembled the feathers as raw materials. After many studies on the feather arrangement of bird wings *(21,22)* and after trying different configurations of bird feathers on the wing skeletons, 3 different types of feathers (Fig. 2C) were chosen. Each single piece of feather was carefully preened and cleaned before it was attached to the wing skeleton (Fig. 2A).

To verify that the power generated by the feathered wings was sufficient to make the robot fly, the team designed 3 field test steps. The wings used in the field were transformed from legacy membrane wings to feathered wings step by step, and the impacts arising from each subtle modification on the wings were reflected in the results of each step in the field test.

The robot in step 1 is armed with an inner membrane wing and an outer membrane wing. In step 2, the outer wing is replaced by a feathered wing, and in step 3, both wings are replaced by feathered wings. The wings in each step are fabricated as follows:

**Feathered wing design and fabrication**

**Step 1: Both the inner and outer wings are membrane wings.** The wing skeletons are covered with flexible and light EPP films (Fig. 2F. 1). To ensure that the inner wing generated sufficient lift while minimizing the weight, the team chose a 3 mm thick, high foaming rate (45 times) thin EPP film to cover the inner wing. Because the outer wing generates thrust and requires more flexibility during flapping movements, a 3 mm thick EPP film with a low foaming rate (30 times) is chosen. A low foaming rate results in more flexibility and more weight.

**Step 2: The inner wing is a membrane wing, and the outer wing is a feathered wing.** The wing skeletons of both wing sections and the membrane covering materials on the inner wing remain unchanged. The supporting beams of the outer wing skeleton are 4 mm hollow carbon fiber rods connected with large outer wing ribs; it is difficult to cover the entire surface with feathers. To solve this issue, part of the outer wing covering the membrane is reserved, as shown in Fig. 2E, as a supporting platform. The wing rib of the outer wing is reserved and unchanged. The reserved membrane (Region R in Fig. 2E) imitates the muscles on a bird's wing tips and facilitates feather attachment on the wing skeleton. It avoids the intentional overlap of the flapping regions of the primary flying feathers. Therefore, the reserved membrane itself does not contribute significant aerodynamic force during flapping movements. Three kinds of selected feathers (Fig. 2C) are glued to the R region of the outer wing to imitate the feather arrangement of a real bird. The long brown feathers serve as the primary flying feathers. The short black feathers resemble the covering feathers of a real bird and are attached to the R region of the outer wing. The final picture of the feathered outer wing is shown in Fig. 2F.2.

**Step 3: Both wings are feathered wings.** After the field test that used WindRider with feathered outer wings was performed, the inner wing was ready to be feathered. As in step 2, the membrane in the G region is reserved to facilitate feather attachment on the skeleton (Fig. 2E). The long brown feathers (Fig. 2C) were glued to the lower half of the G region to serve as secondary flying feathers, while the short black feathers were glued to the upper half of the G region as a secondary covering feathers. The final picture of the inner feathered wing is shown in Fig. 2F. 3.

Fabrication of the feathered wing proved to be very challenging in the sense that the amount of glue on each feather had to be carefully controlled to avoid adding too much weight. The wet region, where a feather is wetted by glue, of every feather had to be precisely controlled to prevent the feathers from losing flexibility due to their subtle feather textures sticking together and turning into rigid boards. WindRider with feathered wings is shown in Movie 1.

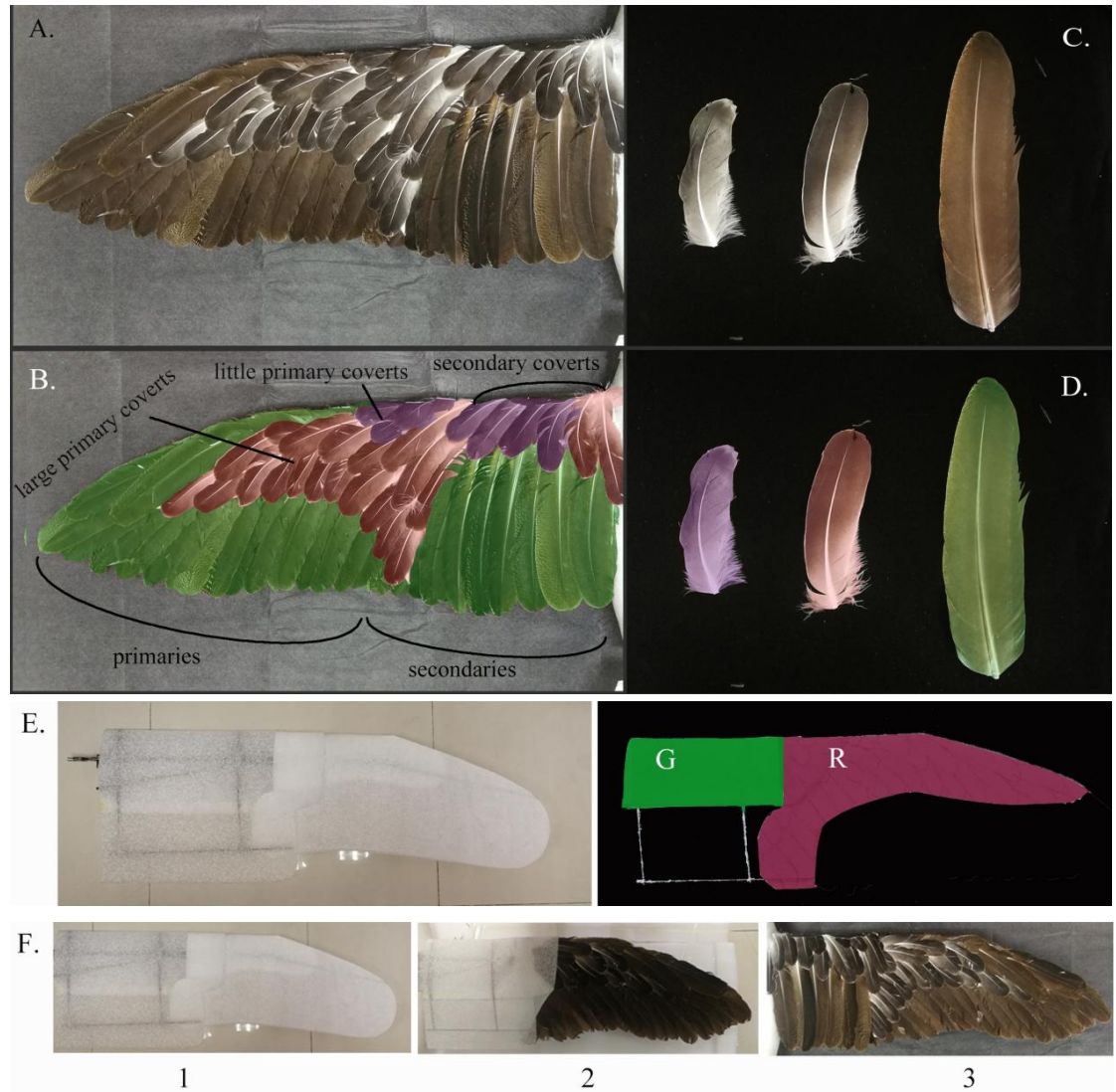

**Fig. 2. Raw feathers, feathered wings and the iterative fabrication process.** (A). Final version of the feathered wings. (B). The feathers in the picture are colored. The original colors are shown in (A) and (C). (C). The raw feathers are categorized into 3 types in terms of the different feather lengths. From left to right: little coverts, large coverts, and primary flying feathers (or secondary flying feathers). (D). The three types of feathers used in (B). Each type of feather is colored in the picture to map the feather to its covering region. (E). The original membrane wing (left picture). The reserved foam film to support feather attachment (right picture). The reserved region in the right picture is colored. (F). The feathered wing fabrication requires 3 iterations, and each iteration gives rise to a new generation of wings. From the previous generation to the new generation requires successful field tests to prove the aerodynamic feasibility. The outcomes of the iteration steps are shown in the pictures: (1) foam film covering the inner and outer wing skeletons; (2) the inner wing with a membrane structure and the outer wing covered in real bird feathers; and (3) both the inner and outer wings covered in authentic bird feathers. The wings shown in pictures (2) and (3) share the same skeletons as the wings in picture (1). The feathers in picture (2) appear to be darker than those in picture (3) because of the light source and background in the picture. They are essentially the same color.

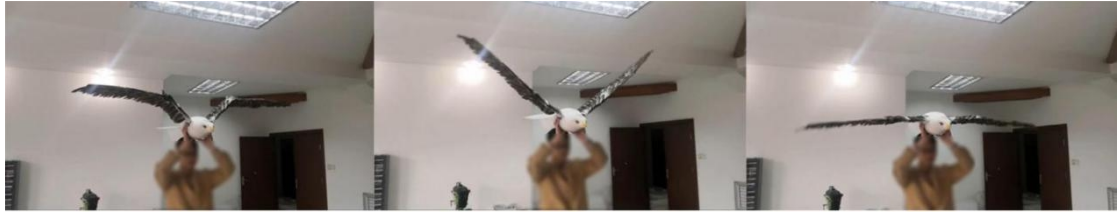

**Movie 1. Three gestures of the feathered wings of WindRider**. Low-frequency flapping (1 Hz) demonstrates the flapping kinematics of WindRider in flight. The face of the person holding the aircraft is intentionally blurred.

## RESULTS

In step 1 of the test, both the inner and outer wings of WindRider were covered with a foam film membrane; WindRider accomplished a successful flight up to 150 s (Movie 2). In step 2 of the test, the outer wings were converted to feathered wings, and the inner wings remained unchanged. The membrane inner wings generated gliding lift, and the outer wings covered with authentic bird feathers provided flapping thrust. On average, the successful flight lasted for 90 s (Movie 3). In step 3 of the test, all wings were covered with authentic bird feathers. The flapping-wing aerial vehicle successfully took off and sustainably flew for up to 63.88 s until it finally landed safely. (Movie 4) proved that the power generated by flapping wings covered in authentic bird feathers is sufficient to make a man-made robot fly.

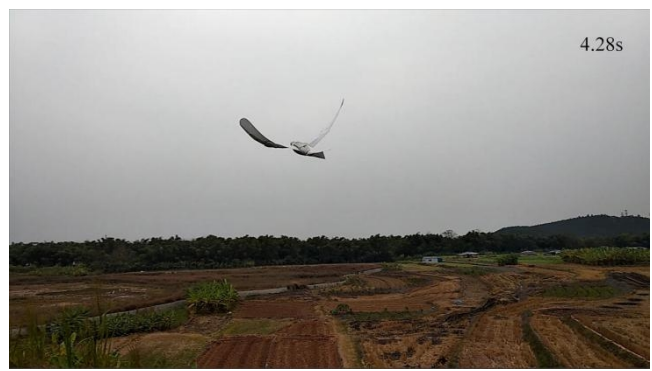

**Movie 2. Field test of WindRider with membrane wings.** Both the inner and outer wings of the robot are covered with a foam film membrane. Its weight is of 588.0 g. The flight duration is 168.88 s. The attitude is controlled by the tail, and landing occurs via unpowered gliding.

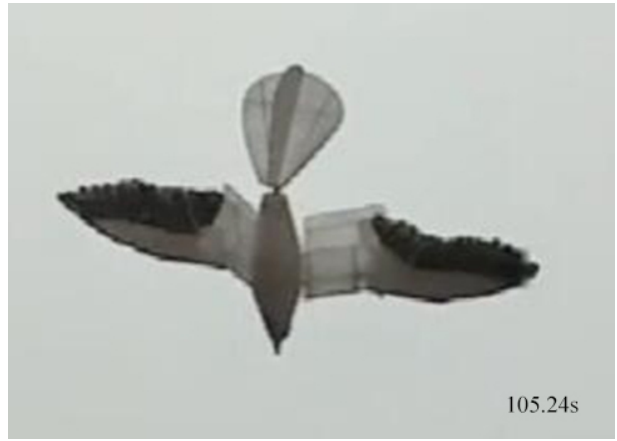

**Movie 3. In step 2 of the field test, the inner wing skeleton is covered with foam film and the outer wing is covered with feathers.** The weight of the whole machine is 630.8 g. The takeoff flapping frequency is approximately 4 Hz. WindRider takes off from an open field and hovers and rises by flapping its wings. Its flight is manipulated by an RC flyer via WFT07. Landing is achieved by reducing the flapping rate until the abdomen touches the ground. The flight duration (from takeoff to landing) is 90.04 s.

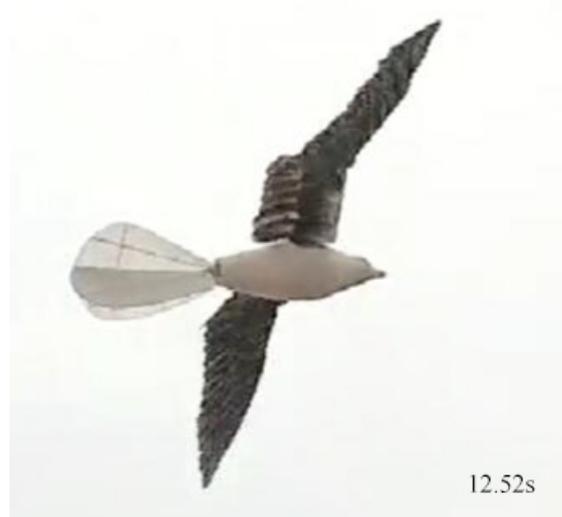

**Movie 4. WindRider in step 3 of the field test; the wings are all covered in authentic bird feathers.** The weight is 667.2 g. The takeoff flapping frequency is ~ 4 Hz. The flight duration (from takeoff to landing) is 63.88 s. The RC flyer controls the aircraft while it hovers in the air; finally, the wing flapping frequency is reduced until WindRider lands safely.

## CONCLUSIONS AND FUTURE WORK

This paper proposes a flapping-wing aerial vehicle whose wings are covered in authentic bird feathers. Different from legacy flapping-wing aerial vehicles, whose wings are membrane structures, the authors replaced the wing membranes with authentic bird feathers. With flapping feathered wings, the robot obtained sufficient lift and thrust in order to take off, fly and land safely. The flight lasted up to 63.88 s. The research demonstrates that flapping wings with carefully arranged bird feathers can generate sufficient aerodynamic force to make a man-made robot fly. In the future, the authors will cover the whole torso and tail with bird feathers to make the robot more similar to a real bird, regardless of the outlook perspective or aerodynamic perspective.

**REFERENCES AND NOTES**

1. J. D. DeLaurier, An ornithopter wing design. *Canadian aeronautics and space journal.* **40**, 10-18 (1994).
2. J. D. DeLaurier, J. M. Harris, A study of mechanical flapping-wing flight. *The Aeronautical Journal.* **97**, 277-286 (1993).
3. M. Karásek, F. T. Muijres, C. De Wagter, B. D. Remes, G. C. de Croon, A tailless aerial robotic flapper reveals that flies use torque coupling in rapid banked turns. *Science* **361**, 1089-1094 (2018).
4. K. Y. Ma, P. Chirarattananon, S. B. Fuller, R. J. Wood, Controlled flight of a biologically inspired, insect-scale robot. *Science* **340**, 603-607 (2013).
5. R. J. Wood, The first takeoff of a biologically inspired at-scale robotic insect. *IEEE transactions on robotics.* **24**, 341-347 (2008).
6. G. C. H. E. De Croon, K. M. E. De Clercq, R. Ruijsink, B. Remes, C. De Wagter, Design, aerodynamics, and vision-based control of the DelFly. *International Journal of Micro Air Vehicles.* **1**, 71-97 (2009).
7. C. J. Pennycuick, *Modelling the Flying Bird* (Elsevier Science, 2008).
8. D. Lentink, U. K. Müller, E. J. Stamhuis, R. de Kat, W. van Gestel, L. L. Veldhuis, P. Henningsson, A. Hedenström, J. J. Videler, J. L. van Leeuwen, How swifts control their glide performance with morphing wings. *Nature* **446**, 1082–1085 (2007).
9. H. Weimerskirch, C. Bishop, T. Jeanniard-du-Dot, A. Prudor, G. Sachs, Frigate birds track atmospheric conditions over months-long transoceanic flights. *Science* **353**, 74-78 (2016).
10. A. Kinkade, Ornithopter, U.S. Patent US20020173217A1 (2002).
11. R. Tedrake, Z. Jackowski, R. Cory, J. W. Roberts, W. Hoburg, "Learning to fly like a bird." In *14th International symposium on robotics research. Lucerne, Switzerland*. (2009).
12. J. Gerdes, A. Holness, A. Perez-Rosado, L. Roberts, A. Greisinger, E. Barnett, J. Kempny, D. Lingam, C.-H. Yeh, H. A. Bruck, S. K. Gupta, Robo Raven: a flapping-wing air vehicle with highly compliant and independently controlled wings. *Soft Robotics.* **1**, 275-288 (2014).
13. M. F. B. Abas, A. S. B. M. Rafie, H. B. Yusoff, K. A. B. Ahmad, Flapping wing micro-aerial-vehicle: kinematics, membranes, and flapping mechanisms of ornithopter and insect flight. *Chinese Journal of Aeronautics.* **29**, 1159-1177 (2016).
14. R. L. Harmon, "Aerodynamic modeling of a flapping membrane wing using motion tracking experiments", thesis, University of Maryland (2008).
15. W. Send, M. Fischer, K. Jebens, R. Mugrauer, A. Nagarathinam, F. Scharstein, "Artificial hinged-wing bird with active torsion and partially linear kinematics." In Proceeding of *28th Congress of the International Council of the Aeronautical Sciences*. (2012).
16. D. Mackenzie, A flapping of wings. *Science.* **335**, 1430-1433 (2012).
17. "Aerodynamic Lightweight Design with Active Torsion", (http://www.festo.com/net/SupportPortal/Downloads/46270/Brosch_SmartBird_en_8s_RZ_110311_lo.pdf, 2011).
18. EPS EPO Foam Adhesive Glue, Online Available: https://item.taobao.com/item.htm?spm=a230r.1.14.87.678e7e74D4Duhh&id=15442058198.
19. E. Chang, L. Y. Matloff, A. K. Stowers, D. Lentink, Soft biohybrid morphing wings with

feathers underactuated by wrist and finger motion. *Science Robotics*. **5**, (2020).
20. D. D. Chin, L. Y. Matloff, A. K. Stowers, E. R. Tucci, D. Lentink, Inspiration for wing design: How forelimb specialization enables active flight in modern vertebrates. *J. R. Soc. Interface* **14**, 20170240 (2017).
21. T. L. Hieronymus, Flight feather attachment in rock pigeons (Columba livia): Covert feathers and smooth muscle coordinate a morphing wing. *J. Anat.* 229, 631–656 (2016).
22. L. Y. Matloff, E. Chang, T. J. Feo, L. Jeffries, A. K. Stowers, C. Thomson, D. Lentink, How flight feathers stick together to form a continuous morphing wing. *Science* **367**, 293–297 (2020).
23. Feather Fan, Online Available: https://detail.tmall.com/item.htm?id=617827733953.
24. RC Transceivers, Online Available: http://www.wflysz.com/product/67.html.